\documentstyle[prl,aps,amsfonts,preprint,epsfig,floats]{revtex}

\input BoxedEPS
\SetRokickiEPSFSpecial  
\HideDisplacementBoxes
\newcommand{\half}{\frac{1}{2}}
\newcommand{\sgn}{{\rm sgn}}

\tighten

\begin{document}

\title{Fractional and Integer Charges from Levinson's Theorem}

\author{E.~Farhi\footnote[0]{e-mail:  farhi@mit.edu,
graham@pierre.mit.edu, jaffe@mit.edu, weigel@ctp.mit.edu}$^{\rm a}$,
N.~Graham$^{\rm a,b}$,
R.~L.~Jaffe$^{\rm a}$, and
H.~Weigel\footnote{Heisenberg Fellow}$^{\rm a}$}

\address{{~}\\$^{\rm a}$Center for Theoretical Physics, Laboratory for
Nuclear Science	and Department of Physics \\
Massachusetts Institute of Technology, Cambridge, Massachusetts 02139 \\
and \\
$^{\rm b}$Dragon Systems, Inc.
Newton, MA 02460\\
{~}
{\rm MIT-CTP\#3004 \qquad hep-th/0007189}}

\maketitle

\begin{abstract}

We compute fractional and integer fermion quantum numbers of static
background field configurations using phase shifts and Levinson's
theorem.  By extending fermionic scattering theory to arbitrary
dimensions, we implement dimensional regularization in a
$1+1$~dimensional gauge theory.  We demonstrate that this
regularization procedure automatically eliminates the anomaly in the
vector current that a naive regulator would produce.  We also apply
these techniques to bag models in one and three dimensions.

\end{abstract}

\section{Introduction}

Many field theory solitons have especially interesting properties when they
are coupled to fermions, because they act as strong background fields that
can drastically alter the Dirac spectrum.  Solitons that break $C$ and $CP$
invariance can introduce asymmetries in the Dirac spectrum, causing the
soliton to carry fermion number.  By adiabatically turning on the soliton
from a trivial background, one can observe this fermion number as a level
crossing in the Dirac spectrum.  In addition, solitons with nontrivial
topological boundary conditions can carry fractional fermion number
\cite{JR,GoldWil,GJ}.

Blankenbecler and Boyanovsky \cite{BB} showed how to use a phase shift
representation of the density of states to calculate these fermion quantum
numbers.  There is no need to consider any interpolation of the background
field.  Together with the bound states, the behavior of the phase shift at
threshold gives the integer charge.  Fractional charges naturally
appear in the behavior of the phase shift at large $k$.

In recent work on variational computations of soliton energies
\cite{us,Kiers}, it has been essential to know these fermion numbers, since
one wants to compare configurations with the same fermion charge.  This work
has led us to extend the method of Blankenbecler and Boyanovsky in two
directions.  First, we have shown how to extend dimensional regularization
to a phase shift formalism.  We are then able to reconsider the anomaly
calculation of Ref.~\cite{BB} starting with an explicitly gauge-invariant
regulator, rather than enforcing gauge invariance by hand on the result. 
We can thus explicitly see the difference between the naive regulator and
the gauge-invariant regulator.  Second, we demonstrate an extension of this
work to chiral bag models in three spatial dimensions, giving a practical
example of a fractional fermion number.

Section II describes the Green's function formalism that relates the phase
shifts to the Fock space expansion for the charge.  This result can then be
strengthened by the use of Levinson's theorem.  In Section III we
demonstrate the effects of the choice of regulator in a one-dimensional
gauge theory.  We show that by choosing a gauge-invariant regulator we
automatically ensure that there are no anomalies in the vector current.  In
Section IV we perform explicit calculations in chiral bag models.  In
Appendix A, we develop the formalism for computing fermion phase shifts in
arbitrary dimensions, and demonstrate that it leads to results that agree
with the standard result from Feynman perturbation theory for the
contribution of the tadpole graph to the effective energy.  In Appendix B,
we derive results used in Section IV for the three-dimensional bag.

\section{Spectral Analysis in Soliton Backgrounds}

\subsection{Density of states}

Our primary tool will be the scattering phase shifts.  The phase
shifts are useful because they unambiguously and quantitatively track
the changes of the spectrum of small oscillations around the
background fields, even though that spectrum is continuous and
infinite, and integrals over it of physical quantities may be divergent.

To derive our results, we consider a single Dirac fermion in $n$ space
dimensions.  We will assume that the soliton background has spherical
symmetry, so that the spectrum decomposes into a sum over eigenchannels
$\alpha$.  In one dimension there are just two channels, for even and odd
parity, while in three dimensions the sum will run over parity and over
either total spin or grand spin, also including the degeneracy within each
channel.  We begin by writing the fermion charge density as
\begin{equation}
j^0(x) = \half\langle\Omega| [\Psi^{\dagger}(x),\Psi(x)] |\Omega\rangle
\label{charge}
\end{equation}
where $x$ is an $n$-dimensional vector.  We have been careful to order the
anticommuting fermion fields to maintain charge conjugation invariance: 
$[\bar{\Psi},\Psi]$ is even under ${\cal C}$ and
$[\bar{\Psi},\gamma_5\Psi]$ and $[\bar{\Psi},\gamma^{\mu}\Psi]$ are odd
under ${\cal C}$.

We make the usual Fock decomposition in terms of the eigenstates
$\psi_\alpha^\omega$ of the single-particle Dirac equation,
\begin{equation}
\Psi(x,t=0) =
\sum_\alpha \left(
	\int_0^\infty b_\alpha^\omega \psi_\alpha^\omega(x) d\omega +
	\int_{-\infty}^0 d_\alpha^\omega {}^\dagger\psi_\alpha^\omega(x) d\omega
\right)
\end{equation}
where the integral over $\omega$ includes both continuum and bound
states.  We have normalized the wavefunctions by
\begin{equation}
\int \psi^\omega_\alpha(x)^\dagger \psi^{\omega'}_{\alpha'}(x) \, dx
= \delta(\omega - \omega') \delta(\alpha - \alpha')
\end{equation}
where the delta function of $\omega$ on the right hand side is
interpreted as a Dirac delta function for continuum states and a Kronecker
delta for bound states.  The creation and annihilation operators satisfy
\begin{equation}
\left\{ b_\alpha^\omega,~b_{\alpha^\prime}^{\omega^\prime}{}^\dagger \right\} =
\left\{ d_\alpha^\omega,~d_{\alpha^\prime}^{\omega^\prime}{}^\dagger \right\}
= \delta(\omega - \omega') \delta(\alpha - \alpha')
\end{equation}
with all other anticommutators vanishing.  Thus the charge density becomes 
\begin{equation}
j^0(x) = -\half \sum_\alpha \left( \int_{-\infty}^\infty \sgn(\omega)
 |\psi^\omega_\alpha(x)|^{2} d\omega \right) \, .
\label{charge2}
\end{equation}

We next consider the Green's function for the fermion field
\begin{eqnarray}
G(x,y,t) &=& i \langle \Omega|T(\Psi(x,t) \Psi(y,0)^\dagger) |
\Omega \rangle \cr
&=&
i \sum_\alpha \left(
\int_{-\infty}^0 d\omega e^{i\omega t} \psi^\omega_\alpha(x)
\psi^\omega_\alpha(y)^\dagger \Theta(-t) -
\int_0^\infty d\omega e^{-i\omega t} \psi^\omega_\alpha(x)
\psi^\omega_\alpha(y)^\dagger \Theta(t) \right)
\end{eqnarray}
and its Fourier transform
\begin{equation}
G(x,y,E) =  \sum_\alpha \int_{-\infty}^\infty \frac{d\omega}{2\pi}
\frac{\psi^\omega_\alpha(x) \psi^\omega_\alpha(y)^\dagger}
{E - \omega + i \sgn(\omega) \epsilon}
\end{equation}
whose trace gives the density of states according to
\begin{equation}
\rho(\omega) = \sgn(\omega) \frac{dN(\omega)}{d\omega}
= {\rm Im~Tr} \:\frac{1}{\pi} \int G(x,x,\omega) \, d^nx
\end{equation}
giving as a result
\begin{equation}
\rho(\omega) = \frac{1}{\pi} \sum_\alpha \int d^nx
|\psi^\omega_\alpha(x)|^2.
\end{equation}
for $|\omega| > m$.  (For $|\omega| < m$, the density of states consists of
a sum of delta functions for each bound state.)
A more useful form of this equation is obtained by subtracting the free
density of states $\rho^0(\omega)$, giving
\begin{equation}
\rho(\omega)  - \rho^0(\omega)= \frac{1}{\pi} \sum_\alpha \int d^nx
(|\psi^\omega_\alpha(x)|^2 - 1) \, .
\end{equation}

We can also express the density of states in each channel in terms of
the S-matrix:
\begin{equation}
\rho_\alpha(\omega) - \rho^0_\alpha(\omega) =
\sgn(\omega) \frac{1}{2\pi i} \frac{d}{d\omega} \log S_\alpha(\omega)
= \sgn(\omega) \frac{1}{\pi} \frac{d\delta_\alpha(\omega)}{d\omega} \, .
\end{equation}
The total density of states is then
\begin{equation}
\rho(\omega) = \sum_\alpha \rho_\alpha(\omega).
\end{equation}

Putting these results into eq.~(\ref{charge2}) we thus obtain an integral
over the continuum and a sum over bound states
\begin{eqnarray}
Q &=& -\sum_\alpha \left(
\int_m^\infty \frac{d\omega}{2\pi} \frac{d\delta_\alpha(\omega)}{d\omega}
+ \sum_{\omega_\alpha^j > 0} \half
-\int^{-\infty}_{-m} \frac{d\omega}{2\pi}
\frac{d\delta_\alpha(\omega)}{d\omega}
- \sum_{\omega_\alpha^j < 0} \half \right)\cr
&=& \frac{1}{2\pi} \sum_\alpha \left( \delta_\alpha(m) -
\delta_\alpha(\infty) - \pi n^>_\alpha + \pi n^<_\alpha -
\delta_\alpha(-m) + \delta_\alpha(-\infty)
\right)
\label{main}
\end{eqnarray}
where $n^>_\alpha$ and $n^<_\alpha$ give the number of bound
states with positive and negative energy respectively in each channel. 
This is our main result.  We will see that it generalizes unchanged to
cases with fractional charges, which appear in the phase shift at
$\omega=\pm\infty$.

\subsection{Levinson's Theorem}

We have seen that the phase shifts keep track of the rearrangement of the
fermion spectrum in a quantitative way, by telling us the density of
states.  Levinson's theorem shows that the phase shifts also
track how many states have left the continuum at each threshold.  This
property will give us a physical motivation for eq.~(\ref{main}), and will
also enable us to simplify it.

Levinson's theorem in three dimensions and in the negative parity channel
in one dimension gives the number of states $N$ that have left the
continuum by passing through the threshold at $m$ as
\begin{equation}
\delta(m) - \delta(\infty) = N\pi.
\end{equation}
These states typically appear as bound states (which give delta
functions in the density of states), though it is possible
that in cases where the spectrum is not charge conjugation invariant,
they can reenter the continuum of states with opposite energy.
In the positive parity channel in one dimension, Levinson's theorem is
modified to
\begin{equation}
\delta(m) - \delta(\infty) = (N-\half) \pi.
\label{levsym}
\end{equation}
In one dimension and in the $\ell=0$ channel in three dimensions,
there is the possibility of a state whose wavefunction goes to a
constant at $r=\infty$, rather than decaying exponentially like a
bound state or oscillating like a continuum state.  Such
``half-bound'' states are on the verge of becoming bound and count as
$\half$ in Levinson's theorem.  In the free case, for example,
$\delta(\omega)=0$ for all $\omega$, but eq.~(\ref{levsym}) still
holds because there is a half-bound state with $\psi = {\rm
constant}$.  For fuller discussion of these results, see Ref.~\cite{Levinson}.

Computing the fermion number of a field configuration now becomes a matter
of simple counting.  We consider each channel separately.  If a bound state
leaves the positive continuum but appears as a positive energy bound state,
it has not changed the fermion number of the configuration.  However, if it
crosses $\omega = 0$ and becomes a negative energy bound state, it is  now
filled in the vacuum and gives a fermion number of one.  Thus the fermion
number of a field configuration is given by
\begin{equation}
Q = \frac{1}{2\pi} \sum_\alpha \left( \delta_\alpha(m) -
\delta_\alpha(\infty) - \pi n^>_\alpha + \pi n^<_\alpha -
\delta_\alpha(-m) + \delta_\alpha(-\infty)
\right) 
\end{equation}
which is exactly (\ref{main}).

This interpretation suggests a further simplification:  since Levinson's
theorem tracks all states that enter and leave the two continua, even in the
presence of CP-violation, we have the restriction that
\begin{equation}
\delta_\alpha(m) - \delta_\alpha(\infty) + \delta_\alpha(-m)
- \delta_\alpha(-\infty) - \pi n_\alpha^< - \pi n_\alpha^> = 0\, ,
\label{consistency}
\end{equation}
so that
\begin{eqnarray}
Q &=& \sum_\alpha \frac{1}{\pi}\left( \delta_\alpha(m) -
\delta_\alpha(\infty) - \pi n_\alpha^> \right) \cr &=& \sum_\alpha
\frac{1}{\pi}\left( \pi n_\alpha^< - \delta_\alpha(-m) +
\delta_\alpha(-\infty) \right).
\end{eqnarray}

In the positive parity channel in one space dimension, we must subtract 1 from
the left-hand side of eq.~(\ref{consistency}) and $\half$ from the
subsequent expressions for $Q$ because of the modification to Levinson's
theorem in eq.~(\ref{levsym}).

\section{Electrostatics and the need for regularization}

The conserved charges we consider are not renormalized.  That is, they do
not receive any contributions from the counterterms of the theory. 
Nonetheless, it is essential to include the effects of the regularization
procedure used to define the theory.  The example of QED in 1+1~dimensions
provides a clear illustration of this subtlety.  Although the theory is
finite, the regularization process is nontrivial.

The Lagrangian is
\begin{equation} {\cal L} = -\frac{1}{4e^2} F_{\mu\nu} F^{\mu\nu} + \half
\left[\bar \Psi, \left(\gamma^\mu (i\partial_\mu - A_\mu) - m\right)\Psi
\right]
\end{equation}
where we again have used the commutator to ensure that the free theory is
$C$ and $CP$ invariant.  In ordinary perturbation theory, the vacuum
polarization diagram computed in $d$ space-time dimensions is
\begin{equation}
\Pi_{\mu\nu}(p) = 2ie^2N_d \int_0^1\hspace{-0.15cm} d\xi \hspace{-0.1cm}
\int \frac{d^dk}{(2\pi)^d}
\frac{ 2\xi(1-\xi) (g_{\mu\nu}p^2 - p_\mu p_\nu) +
g_{\mu\nu} \left(m^2-p^2 \xi(1-\xi) + k^2 (\frac{2}{d} - 1)\right) }
{(k^2 + p^2 \xi(1-\xi) - m^2)^2}
\label{poltensor}
\end{equation}
where $2 N_d$ is the dimension of the Dirac algebra.   If we had not
regulated the theory by analytically continuing the space-time dimension,
we would not have found the last term, which vanishes if we set $d=2$ from
the outset.  Keeping $d\ne 2$ shows that this term exactly cancels the
two terms that precede it, leaving the transverse form of the vacuum
polarization that is required by gauge invariance.  Thus we must
include in our definition of the field theory the additional
information that the theory is regulated in order to preserve gauge
invariance at the quantum level, and dimensional regularization
provides a convenient way to implement this requirement.

The vacuum polarization diagram reflects the effect of the anomaly.   The
anomaly is obtained from the leading correction to the vector current,
which is related to the polarization tensor by
\begin{equation}
j_\mu(x)=\int d^dy\Pi_{\mu\nu}(x-y)A^\nu(y)
\end{equation}
where $\Pi_{\mu\nu}(x)$ denotes the Fourier transform of
eq.~(\ref{poltensor}).  Thus a completely transverse polarization tensor
corresponds to a conserved vector current. Setting $d=2$ from the outset
gives an anomalous vector current
\begin{eqnarray}
\partial_\mu j^\mu &=&
\partial_\mu \bar\Psi\gamma^\mu\Psi =
\frac{e N_d}{\pi} \partial_\mu A^\mu \cr
\partial_\mu j^\mu_5 &=&
\partial_\mu \bar\Psi\gamma^\mu \gamma_5 \Psi = 0 \,.
\label{naiveanomaly}
\end{eqnarray}
Including the contribution proportional to $\frac{d}{2}-1$
in eq.~(\ref{poltensor}) transfers this anomaly to the axial current
\begin{eqnarray}
\partial_\mu j^\mu &=&
\partial_\mu \bar\Psi\gamma^\mu\Psi = 0 \cr
\partial_\mu j^\mu_5 &=&
\partial_\mu \bar\Psi\gamma^\mu \gamma_5 \Psi =
\frac{e N_d}{2 \pi} \epsilon^{\mu\nu} F_{\mu\nu} \, .
\end{eqnarray}

If we choose a configuration with $A_1 = 0$ and adiabatically turn on a
configuration $A_0(x)$ between $t=-\infty$ and $t=0$, then integrating
eq.~(\ref{naiveanomaly}) we obtain
\begin{equation}
Q = \int j^0(x) dx = \frac{e N_d}{2\pi} \int A_0(x) dx
\label{naivecharge}
\end{equation}
for the naive regulator while $Q = 0$ in dimensional regularization.

The phase shift approach shows exactly the same behavior.   We
consider the example of an electrostatic square well potential with
depth $\varphi$ and width $2L$.  First we ignore subtleties of
regularization and compute directly in $d=2$.  The phase shift in the
negative parity channel $\delta^-(\omega)$ is determined by
\begin{equation}
\frac{m+\omega}{k} \tan(kL + \delta^-(\omega)) = \frac{m+\omega+e\varphi}{q}
\tan qL
\end{equation}
with $k = \sqrt{\omega^2 - m^2}$ and  $q =
\sqrt{(\omega+e\varphi)^2 - m^2}$.  Similarly, the phase shift in the
positive parity channel $\delta^+(\omega)$ is determined by
\begin{equation}
\frac{k}{m+\omega} \tan(kL + \delta^+(\omega)) = \frac{q}{m+\omega+e\varphi}
\tan qL \, .
\end{equation}
As $\omega\to\pm\infty$, the total phase shift approaches
$\pm 2 e\varphi L$, giving a fractional contribution to the total fermion
charge in agreement with eq.~(\ref{naivecharge}).  Although such
fractional charges are possible (and we will see examples of them in the
next section), in this case the result indicates that the method of
calculation has not preserved gauge invariance.

To preserve gauge invariance, we regularize the computation by computing
the phase shifts as analytic functions of the space dimension $n$.  Since we
are only concerned with the contribution from $\omega \to\pm\infty$, we can
consider just the leading Born approximation.  In Appendix A we have
extended the method of dimensional regularization of the phase shifts in
Ref.~\cite{us} to fermions.  In arbitrary dimensions, the total phase
shift is a sum over channels labeled by total spin $j =
\frac{1}{2},\frac{3}{2}\dots$ and by parity. Summing over parity, the
leading Born approximation to total phase shift in each $j$ channel is
given by
\begin{equation}
\delta^{(1)}_{n,j} = \omega e \pi \int_0^\infty A_0(r)
\left(J_{\frac{n}{2} + j - \frac{3}{2}}(kr)^2 +
J_{\frac{n}{2} + j - \frac{1}{2}}(kr)^2 \right) r dr
\end{equation}
which has degeneracy $d(j)$ given by eq.~(\ref{degfact}).  Summing
over $j$ using eq.~(\ref{cond2}) and eq.~(\ref{BesselId}) yields the
leading Born approximation to the total phase shift in $n$ space
dimensions
\begin{eqnarray}
\delta^{(1)}_n(\omega) &=&
\omega e \pi N_d \sum_{\ell = 0}^{\infty} D(\ell) \int_0^\infty A_0(r)
J_{\frac{n}{2}+\ell-1}(kr)^2 r dr \cr
&=& \omega k^{n-2} \frac{N_d e \pi}{2^{n-2} \Gamma(\frac{n}{2})^2}
\int_0^\infty A_0(r) r^{n-1} dr \cr
&=& \omega k^{n-2} \frac{N_d L^n e\varphi\pi}{2^{n-2} n 
\Gamma(\frac{n}{2})^2}
\end{eqnarray}
which reduces to $N_d e\varphi L$ if we send $n\to 1$ and take the
limit as $\omega\to\pm\infty$.  But the order of these limits is essential: 
if we first regulate the theory by holding the dimension fixed at $n<1$, we
then see that the contribution as $\omega\to\pm\infty$ vanishes.  Only
after we have taken the $\omega\to\pm\infty$ limits do we send $n\to 1$.
This procedure, dictated by dimensional regularization, preserves gauge
invariance and gives no fractional charge.

We note that other regularization methods commonly used in phase shift
calculations, such as zeta-function regularization, would not preserve
gauge invariance and would thus lead to the same spurious fractional
result.

\section{Fractional Charges}

\subsection{Chiral bag model in one dimension}

Chiral bag models provide simple illustrations of fractional fermion
numbers. We begin with a Dirac fermion in one dimension on the half-line
$x>0$, subject to the boundary condition
\begin{equation}
ie^{i\gamma_5 \theta} \Psi = \gamma^1 \Psi
\label{boundary1d}
\end{equation}
at $x = 0$, with  $-\frac{\pi}{2} \leq \theta \leq
\frac{\pi}{2}$.  We consider the Dirac Hamiltonian
\begin{equation}
\gamma^0(-i\gamma^1\partial_x + m) \Psi = \omega \Psi
\end{equation}
in the basis $\gamma^0 = \sigma_3$ and $\gamma^1 = i\sigma_2$.
The free solutions are
\begin{equation}
\phi^\pm(x) = \left( \matrix{ \pm k \cr \omega - m} \right) e^{\pm ikx}
\end{equation}
where $k = \sqrt{\omega^2 - m^2}$.  The full solutions are
then given in terms of the phase shifts by
\begin{equation}
\psi^\pm(x) = \phi^-(x) + e^{2i\delta^\pm(\omega)} \phi^+(x).
\end{equation}
We can then solve for the phase shifts using the boundary condition,
eq.~(\ref{boundary1d}).  We obtain
\begin{eqnarray}
\cot \delta^+(\omega) &=& -\frac{k}{\omega - m} \tan \beta \cr
\tan \delta^-(\omega) &=& \frac{k}{\omega - m} \tan \beta
\end{eqnarray}
where $\beta = \frac{\pi}{4}  - \frac{\theta}{2}$.  To find
the bound states, we look for solutions of the form
\begin{equation}
\phi(x) = \left( \matrix{ i\kappa \cr \omega - m} \right) e^{-\kappa x}
\end{equation}
with $\kappa = \sqrt{m^2 - \omega^2}$.  Imposing the boundary condition
gives
\begin{equation}
\kappa = (m+\omega) \tan \beta
\end{equation}
so there is always exactly one bound states for $-\frac{\pi}{2} \leq
\theta \leq \frac{\pi}{2}$.  Plugging these results into eq.~(\ref{main}),
we find that the fermion number is $\frac{\theta}{\pi}$.

\subsection{Chiral bag model in 3 dimension}

This simple model generalizes naturally to 3 dimensions.  We consider an
isodoublet of Dirac fermions subject to the boundary condition
\begin{equation}
ie^{i\theta \vec\tau \cdot \hat n \gamma_5} \Psi
 = \vec \gamma \cdot \hat n \Psi
\label{bagbound}
\end{equation}
imposed on a sphere of radius $R$, where $\vec\tau$ are the isospin Pauli
matrices.  This condition is not invariant under space and isospin
rotations individually, but it is invariant under combined space and
isospin rotations, and under parity.  Thus we can decompose the scattering
problem into eigenchannels labelled by grand spin $G = 0,~1,~2\dots$ and
parity. This calculation is outlined in Appendix B.

In the $G=0$ channel, the phase shifts in the two parity channels are
\begin{equation}
\delta_0^\pm(\omega) =
{\rm Arg} \left( ih_1(kR) \frac{k}{\pm \omega-m} \cos \theta
+ ih_0(kR) (1 \mp \sin\theta) \right)
\label{bag0}
\end{equation}
where $h_n(x)$ are spherical Hankel functions of the first kind and
$k = \sqrt{\omega^2 - m^2}$.  Extracting the contribution from the
$G=0$ channel is straightforward.  Using eq.~(\ref{main}), we find a
contribution of $-\frac{\theta}{\pi}$ to the fermion number.

For $G>0$ there are two states for each choice of $G$ and parity.  The total
phase shift in each channel is given by
\begin{eqnarray}
\delta_G^\pm(\omega) = {\rm Arg} && \left(
\sin\theta \frac{h_G(kR)^2}{(\omega \mp m)R}
+ \frac{k}{\pm\omega-m} h_G(k R) \left(h_{G+1}(kR) - h_{G-1}(kR)\right)
\phantom{\left[\left(\frac{1}{1}\right)^2\right]} \right. \cr
&& \left.
- \cos^2 \theta \left[h_{G+1}(kR)
h_{G-1}(kR) \left(\frac{k}{\pm\omega-m}\right)^2 - h_G(kR)^2 \right]
\right) \, .
\label{bag1}
\end{eqnarray}
The contribution from $G>0$ needs to be treated with care.  For large $k$,
the dominant contribution comes from $G+\half \approx kR$.  We must follow a
consistent regularization procedure in order to obtain the correct order of
limits. In this model a simple cutoff suffices.  We first compute the total
phase shift at $k = \Lambda$, summed over all partial waves.  This sum then
has a smooth limit as $\Lambda \to\infty$.  If we had taken the limit in
the other order by considering the $k\to\infty$ limit in each partial wave
separately, we would incorrectly conclude that the contribution from $G>0$
was identically zero.

\begin{figure}[hbt]
\centerline{\BoxedEPSF{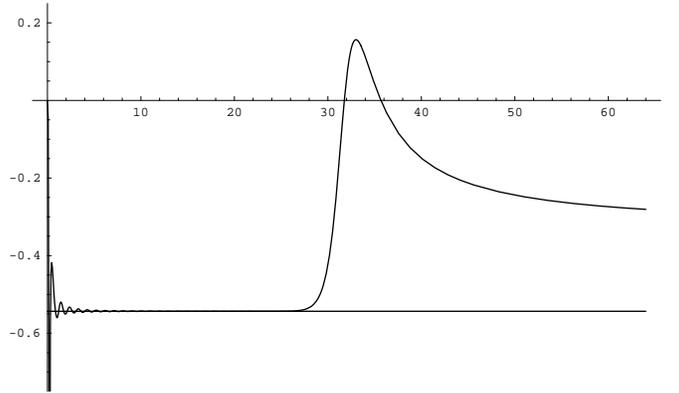 scaled 500}}\bigskip
\caption{\sl Numerical computation of
$\frac{d}{d\theta} \sum_{G=0}^{G_{\rm max}} (2 G + 1)
\left(\delta_G^+(\omega) + \delta_G^-(\omega)
- \delta_G^+(-\omega) - \delta_G^-(-\omega) \right)$,
evaluated at $\theta = \frac{3\pi}{8}$ and $m = 0$, summed up to
$G_{\rm max} = 32$, and plotted a function of $z = \omega R$ for $\omega > 0$.
The larger values of $G$ that we have omitted are negligible for
$z\ll G_{\rm max}$.  For $z\approx G_{\rm max}/2$ we see good agreement with
the result of Ref.~\protect\cite{GJ}, which is shown by a horizontal line.
For $z\gg G_{\rm max}$ we see only the contribution from $G=0$, which would
lead to an incorrect result for the charge.
}
\label{3dbag}
\end{figure}

For numerical computations, it is convenient to first consider
$\frac{dQ}{d\theta}$ and then integrate to obtain $Q$ as a function of
$\theta$.  Figure~\ref{3dbag} shows the $\frac{dQ}{d\theta}$ computation
for one value of $\theta$.  Fixing a large value of the cutoff $\Lambda$,
we then sum over both parities from $G=0$ up to $G_{\rm max} \gg \Lambda
R$, so that the contribution from the higher values of $G$ is negligible. 
We then compute the $\frac{dQ}{d\theta}$ for each $\theta$ from
eq.~(\ref{main}) using $k=\Lambda$ in place of $k = \infty$.

Carrying out this computation numerically for each $\theta$,
we find agreement with the established result \cite{GJ}
\begin{equation}
\frac{dQ}{d\theta} = -\frac{1}{\pi}(1 - \cos 2 \theta).
\end{equation}
so that
\begin{equation}
Q = -\frac{1}{\pi}(\theta - \sin\theta \cos\theta).
\end{equation}

\section{Conclusions}

By explicitly introducing appropriate regulators, we have been able to
unambiguously compute fermion quantum numbers in a variety of models. We
have seen that by introducing the regulator as part of the definition of the
theory, we have no need (and no freedom) to adjust the results of our
computations after the fact.  This determinism leads to a computation that 
is both theoretically elegant and computationally practical.

\subsection*{Acknowledgments}
We would like to thank S.~Forte, J.~Goldstone, and V.~P.~Nair for
helpful conversations, suggestions, and references.  This work is supported
in part by funds provided by the U.S. Department of Energy (D.O.E.) under
cooperative research agreement \#DF-FC02-94ER40818 and the Deutsche
Forschungsgemeinschaft (DFG) under contract We 1254/3-1.
\appendix

\section{Dirac Equation in Arbitrary Dimensions}

In this Appendix we discuss the Dirac equation for a radially symmetric
potential $V=V(r)$ for an arbitrary number of spatial dimensions $n$.  We
calculate the leading Born approximation to the phase shift in $n$
dimensions and find its contribution to the Casimir energy.  We verify
that this result agrees with what we obtain from the tadpole graph in
standard Feynman perturbation theory using dimensional regularization.

Our starting point is the free Dirac equation in $n$ spatial dimensions,
\begin{equation}
\left(-i\vec{\alpha}\cdot\vec{\partial} + \beta m\right)\Psi
=H\Psi=\omega\Psi \, .
\label{freedir}
\end{equation}
The spinor $\Psi$ has $2N_d$ components and accordingly the Dirac 
matrices $\alpha_j,j=1,\ldots,n$ and $\beta$ have $2N_d\times2N_d$
elements, where $N_d=2^{(n-1)/2}$ for $n$ odd and $N_d=2^{n/2}$ for $n$ even.
We will generalize the case of $n$ odd, though our results will not
depend on this choice.  We choose the basis
\begin{equation}
\beta=\pmatrix{1 & 0 \cr 0 & -1}
\qquad{\rm and}\qquad
\alpha_j = \pmatrix{0 & \Lambda_j \cr \Lambda_j & 0} \qquad
j=1,\ldots,n\, .
\label{dirmat}
\end{equation}
The Clifford algebra is obtained by demanding the anti--commutator
$\{\Lambda_i,\Lambda_j\}=2\delta_{ij}$. In analogy to the 
spin generators in $n=3$ we define the commutator
\begin{equation}
[\Lambda_i,\Lambda_j]=2i\Sigma_{ij}
\label{spin}
\end{equation}
which obeys the $SO(n)$ commutation relation
\begin{equation}
[\Sigma_{ij},\Sigma_{kl}]=
i\left(\delta_{ik}\Sigma_{jl}+\delta_{jl}\Sigma_{ik}
-\delta_{il}\Sigma_{jk}-\delta_{jk}\Sigma_{il}\right)\, .
\label{sigcom}
\end{equation}
We define the orbital angular momentum operator
\begin{equation}
L_{ij}=-i\left(x_i\partial_j-x_j\partial_i\right)
\label{orbit}
\end{equation}
which also satisfies the $SO(n)$ algebra
\begin{equation}
[L_{ij},L_{kl}]=
i\left(\delta_{ik}L_{jl}+\delta_{jl}L_{ik}
-\delta_{il}L_{jk}-\delta_{jk}L_{il}\right)\, .
\label{orbitcom}
\end{equation}
We can then put these together to form the total spin operator
\begin{equation}
J_{ij}=L_{ij}+\frac{1}{2}\Sigma_{ij}
\label{totspin}
\end{equation}
which commutes with the Hamiltonian: $[H,J_{ij}]=0$.  Having obtained the
algebra, we next need to find the Casimir eigenvalues of 
\begin{equation}
L^2=\frac{1}{2}\sum_{i,j}L_{ij}^2 \quad
\Sigma^2=\frac{1}{2}\sum_{i,j}\Sigma_{ij}^2\quad{\rm and}\quad
J^2=\frac{1}{2}\sum_{i,j}J_{ij}^2 \, .
\label{casimir}
\end{equation}
The eigenvalues of $L^2$ are those of $SO(n)$, $\ell(\ell+n-2)$.  To find
$\Sigma^2$, we consider its trace, which is just the number of independent
matrices $\Sigma_{ij}$, $\half n (n-1)$.  Then to obtain the Casimir
eigenvalue of $J^2$, all we need to find is $\langle L\cdot \Sigma\rangle =
\langle \frac{1}{2}\sum_{i,j}L_{ij}\Sigma_{ij} \rangle$.  We use the second
order equations obtained from eq.~(\ref{freedir}), which are generalized
Bessel equations. We first remark that $\hat{r}\cdot\vec{\Lambda}$ has
zero total spin ($[\hat{r}\cdot\vec{\Lambda},J_{ij}]=0$) and eigenvalue
$\ell=1$ with respect to $L^2$. Therefore the appropriate spinor with
definite parity can be parameterized as
\begin{equation}
\Psi=\pmatrix{if(r)Y_{l,s,j}\cr 
g(r)\left(\hat{r}\cdot\vec{\Lambda}\right)Y_{l,s,j}}
\label{spinor}
\end{equation}
where $Y_{l,s,j}$ denote generalized spinor spherical harmonics.
The radial functions obey the coupled first order equations
\begin{eqnarray}
\left[\partial_r+\frac{n-1+R}{2r}\right]g(r)&=&(m-\omega)f(r)
\nonumber \\
\left[\partial_r+\frac{n-1-R}{2r}\right]f(r)&=&(m+\omega)g(r)
\label{first}
\end{eqnarray}
where $R=\langle L\cdot \Sigma\rangle+n-1$ contains the desired 
eigenvalue. We can decouple these equations to obtain second order
equations for $f(r)$ and $g(r)$.  By demanding that $f(r)$ and $g(r)$ 
obey generalized Bessel equations with orbital angular momentum
$\ell$ and $\ell^\prime$ respectively, we find
\begin{equation}
R=1\pm(n-2-2\ell)\qquad {\rm and}\qquad 
R=-1\pm(n-2+2\ell^\prime)\, .
\label{Rell1}
\end{equation}
In view of the above mentioned properties of 
$\hat{r}\cdot\vec{\Lambda}$ we have $\ell^\prime=\ell\pm1$. Hence
the two relations in eq.~(\ref{Rell1}) are consistent if 
\begin{equation}
R=n+2\ell-1\quad{\rm for}\quad \ell^\prime=\ell+1
\qquad{\rm and}\qquad
R=3-n-2\ell\quad{\rm for}\quad \ell^\prime=\ell-1 \,. 
\label{Rell2}
\end{equation}
Putting these results together we find the Casimir eigenvalue
\begin{equation}
J^2=\ell(\ell+n-2)+\frac{n(n-1)}{8}
+\cases{\ell, & $\ell^\prime=\ell+1$\cr
2-n-\ell, & $\ell^\prime=\ell-1$}\, .
\label{Jcas1}
\end{equation}
Defining $j=\frac{1}{2}(\ell+\ell^\prime)$ yields
\begin{equation}
J^2=\left(j-\frac{1}{2}\right)\left(j+n-\frac{3}{2}\right)
+\frac{n(n-1)}{8}
\label{Jcas2}
\end{equation}
for both cases. The above definition of $j$ also ensures
that (as for $n=3$) there are two independent solutions for a given $j$:
(i) $\ell=j+\frac{1}{2},\ell^\prime=j-\frac{1}{2}$ and (ii)
$\ell=j-\frac{1}{2},\ell^\prime=j+\frac{1}{2}$.  These solutions have
opposite parity.

Finally, we have to find the degeneracy factor, $d(j)$.  We use two trace
relations that are simple in the basis appropriate for $\Sigma_{ij}$ and
$L_{ij}$.  Written in the basis for $J_{ij}$ they connect different
representations and provide information about the degeneracy factors.  We
have
\begin{equation}
\sum_{j=\ell-\half}^{j=\ell+\half} d(j)(R(j)-n+1) = 0
\label{cond1}
\end{equation}
and
\begin{equation}
\sum_{j=\ell-\half}^{j=\ell+\half} d(j) = N_d D(\ell) \, .
\label{cond2}
\end{equation}
where we have defined $d(-\half) = 0$ independent of $n$.  The first
condition is nothing but the tracelessness of  $\langle L\cdot
\Sigma\rangle$ while the second gives the number of states for a given
orbital angular momentum $\ell$,
\begin{equation}
D(\ell)=\frac{\Gamma(n+\ell-2)}{\Gamma(n-1)\Gamma(\ell+1)} (n+2\ell-2) \,.
\end{equation}
Eq. (\ref{cond1}) can be re--expressed as 
\begin{equation}
d(\ell+\half)=\frac{n+\ell-2}{\ell}\, d(\ell-\half)
\label{cond11}
\end{equation}
which after substitution into eq.~(\ref{cond2}) yields
the final result
\begin{equation}
d(j)=N_d (j+\half)\frac{\Gamma(n+j-\frac{3}{2})}
{\Gamma(n-1)\Gamma(j+\frac{3}{2})}\,.
\label{degfact}
\end{equation}
Equation (\ref{cond11}) represents a recursion relation between $d(j)$ and
$d(j+1)$ that can straightforwardly be shown to be satisfied by the
degeneracy factor of eq.~(\ref{degfact}). We note that as $n\to 1$,
eq.~(\ref{degfact}) gives zero in all channels except $j =
\half$, where it is one.  As in the bosonic case, this limit gives the
reduction to the positive and negative parity channels in one dimension.

To show how the dimensional regularization of the phase shifts
corresponds to ordinary dimensional regularization of Feynman
perturbation theory, we compute the contribution to the Casimir energy
from the leading Born approximation to the phase shifts using the
method of \cite{us}, and compare it to the energy of the tadpole
graph, with both quantities evaluated in $n$ dimensions.

The leading Born approximation to the phase shift is, summing over parity
channels,
\begin{equation}
\delta^{(1)}_{n,j}(k) = -\frac{\pi}{2} \int_0^\infty dr V(r) r
\left(J_{\frac{n}{2} + j - \frac{3}{2}}(kr)^2 +
J_{\frac{n}{2} + j - \frac{1}{2}}(kr)^2 \right)
\end{equation}
for $k = \sqrt{\omega^2 - m^2}$.  The leading Born approximation to the
Casimir energy is then given by
\begin{equation}
\Delta E_n^{(1)}[\vec\phi\,]= -\frac{1}{\pi} \int_0^\infty dk
(\sqrt{k^2 +m^2}  - m) \sum_j d(j) \frac{d\delta_{n,j}^{(1)}}{dk} \,.
\end{equation}

Summing over all channels we find the total phase shift
\begin{eqnarray}
\delta^{(1)}_{n}(k) &=&
\sum_{j=\half, \frac{3}{2} \dots} d(j) \delta^{(1)}_{n,j}(k) \cr
&=& -\frac{\pi}{2} \int_0^\infty dr V(r) r
\sum_{j=\half, \frac{3}{2} \dots} d(j)
\left(J_{\frac{n}{2} + j - \frac{3}{2}}(kr)^2 +
J_{\frac{n}{2} + j - \frac{1}{2}}(kr)^2 \right) \cr
&=& -\frac{\pi}{2} \int_0^\infty dr V(r) r
 \sum_{\ell=0,1,2 \dots}  \left(
d(\ell + \half) J_{\frac{n}{2} + \ell - 1}(kr)^2 +
d(\ell - \half) J_{\frac{n}{2} + \ell - 1}(kr)^2 \right) \cr
&=& -\frac{\pi}{2} \int_0^\infty dr V(r) r
\sum_{\ell=0,1,2 \dots} N_d D(\ell)
J_{\frac{n}{2} + \ell - 1}(kr)^2  \, .
\end{eqnarray}

Using the Bessel function identity
\begin{equation}
\sum_{\ell=0}^{\infty}
\frac{(2q+2\ell)\Gamma(2q+\ell)}{\Gamma(\ell+1)}J_{q+\ell}(z)^2 =
\frac{\Gamma(2q+1)}{\Gamma(q+1)^2}
\left(\frac{z}{2}\right)^{2q}
\label{BesselId}
\end{equation}
and setting $q=\frac{n}{2} - 1$, we sum over $\ell$ and obtain for the
Casimir energy
\begin{equation}
\Delta E^{(1)}[\vec\phi\,]=
2N_d \frac{\langle V \rangle }{(4\pi)^\frac{n}{2}
\Gamma\left(\frac{n}{2}\right)} (n-2) \int_0^\infty (\omega-m) k^{n-3}\, dk
\end{equation}
where
\begin{equation}
\langle V \rangle = \int V(x) d^nx =
\frac{2\pi^\frac{n}{2}}{\Gamma\left(\frac{n}{2}\right)}
\int_0^\infty V(r) r^{n-1}dr \, .
\end{equation}
The $k$ integral can be calculated in the vicinity of $n=\frac{1}{2}$ and
then analytically continued, yielding
\begin{equation}
\int_0^\infty (\omega - m) k^{n-3} \, dk = - \frac{m^{n-1}}{4 \sqrt{\pi}}
\Gamma\left(\frac{1-n}{2}\right) \Gamma\left(\frac{n-2}{2}\right)\, .
\end{equation}
Hence we find
\begin{equation}
\Delta E^{(1)}[\vec\phi\,]=
-2N_d \frac{\langle V \rangle }{(4\pi)^\frac{n+1}{2}}
\Gamma\left(\frac{1-n}{2}\right) m^{n-1}
\end{equation}
which is exactly what we obtain using standard dimensional regularization
for the tadpole diagram in $n+1$ space--time dimensions.

\section{Chiral Bag Model S-matrix}

In this Appendix we outline the derivation of eq.~(\ref{bag0}) and
eq.~(\ref{bag1}).  We begin with spinors that are eigenstates of parity and
total grand spin \cite{Al96}, where grand spin $\vec G$ is the sum of total
spin $\vec{j}=\vec{l}+\frac{1}{2}\vec{\sigma}$ and isospin $\half
\vec\tau$.  For a given grand spin $G$ with $z$-component $M$, we will find
the scattering wavefunctions in terms of the spherical harmonic functions
${\cal Y}_{j,\ell}$ with $j=G\pm\frac{1}{2}$ and $\ell=j\pm\frac{1}{2}$. 
These are two-component spinors in both spin and isospin space.  While
grand spin is conserved, the boundary condition in eq.~(\ref{bagbound})
mixes states with different ordinary spin $j$.  For the channels with
parity $(-)^G$ we have two spinors that solve the Dirac equation away from
the boundary:
\begin{equation}
\pmatrix{ig_1(r){\cal Y}_{G+\frac{1}{2},G}\cr
f_1(r){\cal Y}_{G+\frac{1}{2},G+1}}\qquad{\rm and}\qquad
\pmatrix{ig_2(r){\cal Y}_{G-\frac{1}{2},G}\cr
f_2(r){\cal Y}_{G-\frac{1}{2},G-1}}\, .
\label{pos1}
\end{equation}
For zero grand spin the second spinor is absent.  We now
introduce linear combinations that define the S-matrix:
\begin{eqnarray}
\Psi_1&=&
\pmatrix{iw^+h_G(kr){\cal Y}_{G+\frac{1}{2},G}\cr
w^-h_{G+1}(kr){\cal Y}_{G+\frac{1}{2},G+1}}
+S^{+,G}_{11}\pmatrix{iw^+h^*_G(kr){\cal Y}_{G+\frac{1}{2},G}\cr
w^-h^*_{G+1}(kr){\cal Y}_{G+\frac{1}{2},G+1}}
\nonumber \\ \nonumber \\ && \hspace{4.8cm}
+S^{+,G}_{21}\pmatrix{iw^+h^*_G(kr){\cal Y}_{G-\frac{1}{2},G}\cr
-w^-h^*_{G-1}(kr){\cal Y}_{G-\frac{1}{2},G-1}}
\label{S11S21} \\ \nonumber \\
\Psi_2&=&
\pmatrix{iw^+h_G(kr){\cal Y}_{G-\frac{1}{2},G}\cr
-w^-h_{G-1}(kr){\cal Y}_{G-\frac{1}{2},G-1}}
+S^{+,G}_{22}\pmatrix{iw^+h^*_G(kr){\cal Y}_{G-\frac{1}{2},G}\cr
-w^-h^*_{G-1}(kr){\cal Y}_{G-\frac{1}{2},G-1}}
\nonumber \\ \nonumber \\ && \hspace{5.1cm}
+S^{+,G}_{12}\pmatrix{iw^+h^*_G(kr){\cal Y}_{G+\frac{1}{2},G}\cr
w^-h^*_{G+1}(kr){\cal Y}_{G+\frac{1}{2},G+1}}
\label{S22S12}
\end{eqnarray}
where $h_\ell(kr)$ refers to the spherical Hankel functions suitable to
parameterize an incoming spherical wave.  We have introduced the
kinematic factors $w^+=\sqrt{1+\frac{m}{\omega}}$ and $w^-=\sgn
(\omega)\sqrt{1-\frac{m}{\omega}}$, where $\omega=\pm\sqrt{k^2+m^2}$
and $m$ are the energy and mass of the Dirac particle respectively.  In the
case $G=0$, the components with $j=G-\frac{1}{2}$ are absent and the
S-matrix has only a single component $S^{+,0}_{11}={\rm
exp}(2i\delta_0^+)$.

Imposing the boundary condition, eq.~(\ref{bagbound}), on these
wavefunctions gives
\begin{eqnarray}
\pmatrix{{\rm cos}\theta & 
i\hat{r}\cdot\vec{\tau}{\rm sin}\theta
-i\hat{r}\cdot\vec{\sigma}\cr
i\hat{r}\cdot\vec{\tau}{\rm sin}\theta
+i\hat{r}\cdot\vec{\sigma}& {\rm cos}\theta}\Psi_n\Bigg|_{r=R}=0
\qquad {\rm for}\qquad j=1,2\, .
\label{scatbound}
\end{eqnarray}
For each $n=1,2$ the projection onto grand spin spherical harmonics
yields two equations, which allows us to extract all four components 
of the S-matrix.  It is convenient to express the result in the form
of a matrix equation:
\begin{eqnarray}
\pmatrix{X&Y\cr \bar{X} &\bar{Y}}
\pmatrix{S^{+,G}_{11} & S^{+,G}_{12}\cr S^{+,G}_{21} & S^{+,G}_{22}}=-
\pmatrix{X^*&Y^*\cr \bar{X}^* &\bar{Y}^*}\equiv -M^*
\label{smatr}
\end{eqnarray}
where the star denotes complex conjugation.  The components of the matrix
$M$ are given by
\begin{eqnarray}
X&=&h^*_{G}(kR){\rm cos}\theta
+\frac{k}{\omega+m}\left(1+\frac{1}{2G+1}{\rm sin}\theta\right)
h^*_{G+1}(kR) \nonumber \\
\bar{X}&=&-h^*_{G}(kR){\rm cos}\theta
+\frac{k}{\omega+m}\left(1+\frac{1}{2G-1}{\rm sin}\theta\right)
h^*_{G-1}(kR) \nonumber \\
Y&=&\frac{2\sqrt{G(G+1)}}{2G+1}\frac{k}{\omega+m}
h^*_{G-1}(kR){\rm sin}\theta \nonumber \\
\bar{Y}&=&\frac{2\sqrt{G(G+1)}}{2G+1}\frac{k}{\omega+m}
h^*_{G+1}(kR){\rm sin}\theta\, .
\label{matrixcomp}
\end{eqnarray}
The total phase shift in this channel is then given by
\begin{equation}
\delta_G^+={\rm Arg}\left({\rm det}S^{+,G}\right)=
{\rm Arg}\left(\frac{{\rm det}M^*}{{\rm det}M}\right)
\label{delpos}
\end{equation}
leading to eq (\ref{bag1}).  For the channels with parity
$(-)^{G+1}$ the computation proceeds analogously from the definition of the
corresponding S-matrix:
\begin{eqnarray}
\Psi_1&=&
\pmatrix{iw^+h_{G+1}(kr){\cal Y}_{G+\frac{1}{2},G+1}\cr
-w^-h_{G}(kr){\cal Y}_{G+\frac{1}{2},G}}
+S^{-,G}_{11}\pmatrix{iw^+h^*_{G+1}(kr){\cal Y}_{G+\frac{1}{2},G+1}\cr
-w^-h^*_{G}(kr){\cal Y}_{G+\frac{1}{2},G}}
\nonumber \\ \nonumber \\ && \hspace{4.9cm}
+S^{-,G}_{21}\pmatrix{iw^+h^*_{G-1}(kr){\cal Y}_{G-\frac{1}{2},{G-1}}\cr
w^-h^*_{G}(kr){\cal Y}_{G-\frac{1}{2},G}}
\\ \nonumber \\ 
\Psi_2&=&
\pmatrix{iw^+h_{G-1}(kr){\cal Y}_{G-\frac{1}{2},{G-1}}\cr
w^-h_{G}{\cal Y}_{G-\frac{1}{2},G}}
+S^{-,G}_{22}\pmatrix{iw^+h^*_{G-1}(kr){\cal Y}_{G-\frac{1}{2},{G-1}}\cr
w^-h^*_{G}{\cal Y}_{G-\frac{1}{2},G}}
\nonumber \\ \nonumber \\ && \hspace{4.9cm}
+S^{-,G}_{12}\pmatrix{iw^+h^*_{G+1}(kr){\cal Y}_{G+\frac{1}{2},{G+1}}\cr
-w^-h^*_{G}(kr){\cal Y}_{G+\frac{1}{2},G}}\, .
\end{eqnarray}

\end{document}